\documentclass[12pt,preprint]{aastex}

\usepackage{natbib}

\def\deg{$^{\rm o}$}

\shortauthors{Fischer et al.}
\shorttitle{Bicone of Mrk~78}

\begin{document}

\title{HST Observations of the Double-Peaked Emission Lines in the Seyfert Galaxy Markarian 78: Mass Outflows from a Single AGN\altaffilmark{1}}

\author{T.C. Fischer\altaffilmark{2},
D.M. Crenshaw\altaffilmark{2},
S.B. Kraemer\altaffilmark{3},
H.R. Schmitt\altaffilmark{4}
R.F. Mushotsky\altaffilmark{5}
J.P. Dunn\altaffilmark{6}}

\altaffiltext{1}{Based on observations made with the NASA/ESA Hubble Space 
Telescope, obtained at the Space Telescope Science Institute, which is 
operated by the Association of Universities for Research in Astronomy,
Inc. under NASA contract NAS 5-26555.}

\altaffiltext{2}{Department of Physics and Astronomy, Georgia State 
University, Astronomy Offices, One Park Place South SE, Suite 700,
Atlanta, GA 30303; fischer@chara.gsu.edu}

\altaffiltext{3}{Institute for Astrophysics and Computational Sciences,
Department of Physics, The Catholic University of America, Washington, DC
20064}

\altaffiltext{4}{Computational Physics, Inc, Springfield, VA 22151-2110;
and Naval Research Laboratory, Washington, DC 20375}

\altaffiltext{5}{Department of Astronomy, University of Maryland, College
Park, MD 20742-2421}

\altaffiltext{6}{Department of Chemistry and Physics, Augusta State University, Augusta, GA 30904}

\begin{abstract}

Previous ground based observations of the Seyfert 2 galaxy Mrk 78 revealed a double set of 
emission lines, similar to those seen in several active galactic nuclei (AGN) from recent surveys. Are the double lines 
due to two AGN with different radial velocities in the same galaxy, or are they due to mass 
outflows from a single AGN? We present a study of the outflowing ionized gas in the resolved 
narrow-line region (NLR) of Mrk 78 using observations from Space Telescope Imaging Spectrograph 
(STIS) and Faint Object Camera (FOC) aboard the {\it Hubble Space Telescope(HST)} as part of 
an ongoing project to determine the kinematics and geometries of AGN 
outflows. From the spectroscopic information, we determined the fundamental geometry of the 
outflow via our kinematics modeling program by recreating radial velocities to fit those seen 
in four different STIS slit positions. We determined that the double emission lines seen in 
ground-based spectra are due to an asymmetric distribution of outflowing gas in the NLR.
By successfully fitting a model for a single AGN to Mrk 78, we show that it is possible to 
explain double emission lines with radial velocity offsets seen in AGN similar to Mrk 78 
without requiring dual supermassive black holes.

\end{abstract}

~~~~~

\clearpage

\section{Introduction}

Narrow line regions of Seyfert galaxies contain photoionized gas due to the nonstellar continuum 
emitted by the AGN and are structured roughly in the shape of a bicone, where the apex of the 
bicone is located in the center of the AGN \citep{Pog88,Sch94}. This likely indicates that the 
ionizing radiation from the nucleus is being collimated by an absorbing material, which could be 
a torus-like structure \citep{Ant93} or perhaps a disk wind \citep{Eli06}. However, the source of 
the NLR gas is not well understood.

An AGN of particular interest is Mrk 78. Classified as an SB galaxy in the NASA/IPAC 
Extragalactic Database (NED), it is also a Seyfert 2 galaxy because only narrow (full width at 
half-maximum [FWHM] $\le 1000$ km s$^{-1}$) emission lines are present in its optical spectra. 
With a redshift of {\it z} = 0.036793 based on Ca T stellar absorption lines \citep{Nel95}, 
Mrk 78 is at a distance of $\sim$ 150 Mpc (for $H_0 = 73$ km s$^{-1}$ Mpc$^{-1}$); at this 
distance, 1$''$ corresponds to a transverse size of $\sim$730 pc. Mrk 78 has been extensively 
studied due to a visual correlation between the its radio source and extended narrow-line region 
(ENLR) \citep{Whi02,Whi04,Whi05}. {\it HST} FOC images of Mrk 78 in [O~III] show that its 
narrow-line region (NLR) consists of two highly luminous regions extending over $\sim 5''$ 
($\sim$ 3.6 kpc) that is in a biconical geometry also seen in several other Seyfert 2 galaxies 
\citep{Sch03}. Thus, Mrk 78 makes for an interesting addition to our 
ongoing study to determine the geometries of the outflows and fueling flows in the NLRs of 
AGN \citep{Das05,Das06,Cre10b,Fis10}. 

A specifically interesting aspect is that ground-based spectroscopic observations of Mrk 78 have 
shown that its narrow emission lines are double peaked \citep{Sar72,Ada73}. Ground-based spectra 
at high spectral resolution depict the [O~III] emission-line profile as two large, well-defined 
peaks separated by $\sim800$ km s$^{-1}$ \citep{Whi88,Nel95}. As of late, there has been 
considerable activity around the study of double emission lines as they have been used as an
indication for the presence of multiple AGN in a single galaxy. As such, Mrk 78 would seem an 
excellent candidate to test for multiple AGN as it is nearby, can be studied in detail, and 
contains dual-peaked lines with a velocity separation similar to those seen in other examples 
where "dual AGN" were decided a more likely scenario than outflows or rotating disks \citep{
Kom08a,Com09,Liu10}.

\citet{Cre10b} showed that velocity offsets between the narrow emission lines and host galaxy 
lines can be explained by the combination of mass outflows and dust extinction in AGN, or even 
an asymmetric distribution of outflowing gas. They also suggested that these explanations could 
possibly work for double-lined AGN, such as Mrk 78, as well. Previous studies had already 
suggested that radial motion via inflows or outflows could be responsible for the narrow line 
asymmetries and velocity shifts \citep{Whi85,Dah88,DeR90}, including long-slit 
ground-based observations that gave evidence for the production of asymmetries via outflows and 
dust obscuration \citep{Sto92,Arr96,Chr97,Alm06,Rod06}. In this paper, we use high resolution 
spectroscopy and imaging to show that the velocity offset seen in Mrk 78 can indeed 
be explained by mass outflows in the NLR.

\section{Observations}

We obtained archival Space Telescope Imaging Spectrograph (STIS) long-slit spectra and Faint 
Object Camera (FOC) images of Mrk 78 from the Multimission Archive at the Space Telescope 
Science Institute (MAST). Four sets of spectra were taken between 1998 February 28 and 1998 
March 01 under Hubble Program ID 7404 (M. Whittle, PI) with a 52$'' \times 0\farcs2$ slit using 
a medium-dispersion G430M grating (4950 - 5240\AA), which includes [O~III] $\lambda$5007. The 
spectral resolution was 0.56 \AA ($\sim30$ km s$^{-1}$ FWHM), ~with an angular resolution of 
$0\farcs051$ per pixel in the cross dispersion direction. Of the four observations, three were 
taken at a position angle of 88\deg ~and the fourth was taken at a position angle of 61\deg. 
Additional details of each observation are given in Tables 1 and 2. 

Figure 1 shows an extended continuum STIS acquisition image of Mrk 78 with additional, 
contoured [O~III] and UV FOC images and positions for the four STIS slits illustrated on the 
[O~III] image. The slits are offset from the optical continuum peak, signified as a $''+''$, 
which was located by aligning the [O~III] and continuum FOC images taken on the same date 
\citep{Sch03} and is also coincident with the infrared peak as observed by \citet{Alm06}. 

The STIS spectra were processed using IDL software developed at NASA's Goddard Space Flight 
Center for the STIS Instrument Definition Team. Cosmic-ray hits were removed before further 
processing. The zero point of the wavelength scales were 
corrected using wavelength-calibration exposures taken after each observation. Finally, the 
spectra were combined, geometrically rectified, and flux calibrated to provide spectral images 
that have a constant wavelength along each column and display fluxes in units of erg s$^{-1}$ 
cm$^{-2}$ \AA$^{-1}$ per cross-dispersion pixel. The calibrated spectral images for each slit 
position can be found in \citet{Whi02}.

We used the STIS spectra to compare our results with those taken from previous ground-based 
spectra, obtained through apertures that included most, if not all, of the NLR in Mrk 78 at 
moderate spectral resolution (80-230 km s$^{-1}$ FWHM) \citep{Vrt85,Nel95}. The STIS datasets consist 
of hundreds of spectra along each slit, allowing us to resolve individual emission-line knots 
that are responsible for a majority of the emission seen in NLRs. To simulate the ground-based 
spectra, we combined all of the STIS spectra along each slit position and then combined the 
spectra from each slit, excluding slit D as it overlaps the other three slits, similar to the 
procedure in \citet{Cre10a}. Figure 2 shows the final co-added spectrum in terms of flux as a 
function of the radial velocity of the [O~III] line, with respect to the systemic velocity 
calculated from our redshift. The significantly separated [O~III] $\lambda$5007 peaks agree 
with those from the moderate-resolution ground-based spectra of \citet{Nel95}, with the 
higher peak being at near systemic velocity ($v_{r} = -50$ km s$^{-1}$) and the secondary red 
peak at $v_{r} = 800$ km s$^{-1}$, meaning any [O~III] emission not covered by the STIS long-slit 
observations does not significantly alter the profile shape or structure. 

To determine the kinematics of the NLR, we incorporated the processed spectral images into a 
program that fits [O~III] $\lambda$5007 emission-line components with Gaussians over an average continuum, 
taken from available line-free regions in the spectra \citep{Das05}. In order to get the maximum 
number of measured lines per data set, spectra were smoothed with a boxcar average using a width 
of 7 pixels. This smoothing allowed for the fitting of extended, low flux [O~III] emission lines, 
without inhibiting us from deblending multiple [O~III] lines dependent on different kinematic 
components along the slit. Calculating the central peak location for each Gaussian allowed us to 
determine the central wavelength for the corresponding [O~III] line, which in turn was used to 
calculate its Doppler shifted velocity. There were three sources of uncertainty in our 
velocity measurements, as detailed by \citet{Das05}. The first is that the measured emission 
lines are not perfect Gaussians, the second comes from emission cloud displacements from the 
center of the $0\farcs2$ slit in the dispersion direction, and the third comes from noisy 
spectra. Errors were converted to velocities and added in quadrature, which produced a total 
maximum error of $\pm$ 60 km s$^{-1}$. Extremely noisy spectra (spectra without detectable 
emission $> 3\sigma$) at locations $\geq 3'' - 3\farcs5$ to the west and $\geq 3\farcs75$ 
($4\farcs25$ for slit D) to the east of the nucleus, were not fitted. Subtracting the systemic 
velocity of the galaxy ($cz =  11037$ km s$^{-1}$) from velocities calculated via [O~III] 
line centroid shifts, we were able to determine the radial velocities along each slit in the 
rest frame of Mrk 78.

\section{Results} 

Figures 3 and 4 show the rest frame radial velocities, FWHM, and normalized fluxes for the 
measured [O~III] $\lambda$5007 emission lines from the four slit positions. Kinematics from the 
lower spectral resolution IR spectrum of \citet{Alm06} and the spectral images of \citet{Whi02} 
agree with our measurements. In general, the structure of the emitting region is 
clumpy in nature, as seen in the large number of flux peaks in each slit that can be matched to 
individual bright knots in the [O~III] image of Figure 2. While each knot appears to have an 
individual peculiar velocity, the velocity field as a whole is relatively well organized. High 
radial velocities on both sides of the nucleus show the distinct asymmetrical red/blue shifts 
characteristic of biconical outflows seen in several other low-redshift Seyfert galaxies 
\citep{Rui01,Das05,Das06,Cre10b,Fis10}. Emission eastward of the optical peak is dominantly 
redshifted from systemic velocity (henceforth `red lobe') while emission westward of the optical 
peak is dominantly blueshifted (henceforth `blue lobe'). Breaking down our simulated ground based 
spectrum of Figure 2 into the individual slits A, B, and C, we find that the redshifted velocity peak 
is primarily due to a specific high-flux knot, best sampled in slit B. Figure 1 contains a circle 
around the location of the redshifted knot and Figure 3 has a bracket in slit B, located over the 
high-flux region which depicts the knot.

\section{Models}

To generate kinematic models, we assume the ionizing radiation responsible for the NLR is 
biconical in nature, as it is the 
simplest geometric shape produced by a central obscuring torus. As with our previous kinematic 
models \citep{Das05,Das06,Cre10b,Fis10}, we also initially assume that the bicone is hollow, 
where the separate NLR emission components represent opposite sides of the bicone. We employed 
our kinematics program from \citep{Das05}, which allows us to recreate the observed radial 
velocities along a fixed slit position by altering various parameters of our model bicone 
outflow.\footnote{Previous studies of Mrk 78 have suggested that the NLR outflow is being 
radially accelerated via radio jets \citep{Whi04,Whi05}. While our model does not include 
any specific source of acceleration, instead simply simulating general outward acceleration, 
we do not detect considerable differences in the velocities or dispersions of emission-line 
knots near radio knots from others in the overall outflow noted in \citet{Whi02}.} 
Our models include seven alterable 
parameters with initial values taken from the [O~III] image in Figure 1 (deprojected height of 
bicone [$z_{max}$], outer half-opening angle [$\theta_{max}$], and position angle) and the kinematics 
from Figures 3 and 4 (maximum velocity [$v_{max}$] and turnover radius [$r_{t}$]), leaving the 
inner half-opening angle ($\theta_{min}$) and inclination of the bicone axis out of the plane of the 
sky to be alterable parameters free of restriction. We adopted our previous velocity law \citep{Fis10}, 
a linear increase starting with zero km s$^{-1}$ at the nucleus and subsequent linear decrease at $r_{t}$ 
ending at zero km s$^{-1}$, because it is the simplest law that matches the observations.

The FOC [O~III] image of Figure 1 shows that the NLR is not symmetric on either side of the 
nucleus. The redshifted, NE cone is less dispersed, more conical in nature, and features cleaner 
kinematics from which we can discern separate kinematic components due to outflow emission in the fore 
and back side of the cone. Conversely, the blueshifted, SW cone appears to have a larger opening angle, with 
floculent, knotty emission and resulting intricate kinematics. Though using initial parameters 
solely from the NE cone was enticing, we found that the bicone did not envelop the entirety of 
the west lobe. Thus, we expanded the model bicone outer opening angle to envelop all NLR emission 
in the [O~III] image in Figure 2. We chose $\theta_{max} = 30$\deg ~as our initial opening 
angle and used the kinematic components of the NE cone as the basis for the majority of our 
fitting. The initial inclination parameter relied on the kinematics seen in Figures 3 and 4, as 
either side was either mostly redshifted or blueshifted, and the imaging in Figure 1 as the 
extended biconical structure on either side of the nucleus would be less pronounced if we 
were to be looking into the bicone \citep{Sch96}. With these guidelines, we determined an 
intermediate angle was most appropriate for a starting position. To find the most successful 
model, we altered parameters across all data sets while applying specific 
centroid offsets and position angles for each slit until a specific set of free parameters 
created an acceptable fit to all four data sets. Final values for all model parameters are 
provided in Table 3. Interestingly, the final opening angle and position angle are nearly 
identical to the apparent, initial parameters taken from Figure 1 ($\theta_{max} = 30$\deg 
~and $P.A. = 70$\deg ~respectively). In comparison to our previous studies \citep{Rui01,Das05,
Das06,Fis10}, Mrk 78 has an extremely large NLR as well as a narrower, thicker (greater 
difference between $\theta_{max}$ and $\theta_{min}$) bicone than all other targets thus far.

Figure 5 displays a good qualitative fit between the model 
created by our final set of parameters and the data. The model matches fairly well to the east of 
the nucleus as kinematics for both sides of the eastern lobe are discernible in multiple slits. 
One discrepancy worth discussing is the lack of model data 
(shaded regions) mainly to the west of the nucleus in Slits B and C between $-0\farcs25$ to 
$0\farcs5$ and $-0\farcs5$ to $1\farcs5$ respectively. As our kinematics model creates a well 
defined geometrical model bicone with a sharp apex, incorporating slit positions that are 
transversely offset from the nucleus will generate models that lack data near the nucleus as 
the slit does not pass over the model bicone. This is due to the simplicity of our model as we 
assume that the bicone has a sharp apex, which is obviously not the case in this and other 
Seyferts \citep{Ant93,Cre10a,Fis10}. Thus, slits B and C lack modeled kinematics in these regions 
as the actual emission lies outside our model bicone geometry.

\section{Discussion}

The available evidence indicates that Mrk 78 is {\it not} a double AGN. Observing the variance in 
flux across the slit, we see many bumps and peaks that are due to large, bright knots of 
emission-line gas, each with their own peculiar velocity inside the general outflow pattern. It 
is one of these knots, a bright red-shifted cloud traveling at $\sim$ 800 km s$^{-1}$ inside of 
the ionizing bicone, that is the source of the offset radial velocity peak. The cloud appears to 
fit well in our modeled outflow pattern, supporting the suggestion by \citet{Cre10a} that uneven 
outflow can account for such dual-peaked profiles, and thus there is no need to assume there is 
a second AGN present. In comparison to our previous studies, kinematics for the NLR in Mrk 78 
actually look rather similar to those in NGC 4151 \citep{Hut98,Cre00a,Das05}, NGC 1068 
\citep{Cre00b,Das06,Das07}, Mrk 3 \citep{Rui01,Cre10b}, and Mrk 573 \citep{Sch09,Fis10}. For 
each of these targets, including Mrk 78, we have successfully employed our biconical outflow 
models to find a good match to the overall flow patterns as a function of distance from the 
central AGN. Though there are some inconsistencies between the model and the data that are 
likely due to the simplified geometry of our bicone model and pecular velocities of individual 
emission-line knots, radial outflow of [O~III] clouds can explain the general, large-scale flow 
pattern of the NLR in Mrk 78, as well as the double peaked emission. 

Does obscuration from dust contribute to the creation of the dual-peaked radial velocity profile 
as mentioned in \citet{Cre10a}? Ground based images of Mrk 78 in B and I give a position angle and 
inclination of the host galaxy as 84\deg ~and 64\deg ~respectively \citep{Sch00,Kin00}. Adding 
this geometry to a geometric bicone model, we can test to see if a smooth distribution of dust 
contributed by the disk could be responsible for the radial velocity profile. Figures 6 and 7 
show simple, three-dimensional geometric models of the ionizing bicone's outer surface using 
$\theta_{max}$, position angle, and inclination values from our final kinematics model and an 
additional two-dimensional galactic disk using the above position angle and inclination values.
Though it is unknown which side of the disk is closer, either geometry in combination with our 
final bicone model results in a radial velocity profile that is either symmetric or asymmetric 
towards the blue. The latter case can be seen in Figure 6 as the redshifted cone should be more 
extincted by dust in the galactic plane. This scenario does not fit our data as the blue-shifted, 
likely less obscured lobe would then contribute a majority of the total emission-line flux. Thus, 
it is unlikely that a dusty disk would be responsible for the given profile. 

There is a prominent dust lane that sweeps over the center of the AGN. Each image in 
Figure 1 shows a distinct band from southeast to northwest across the nucleus with a position 
angle of $\sim$123\deg ~southeast to northwest that has flux much lower than the emission 
surrounding it. More quantitatively, we obtained an additional STIS G750M (5450 - 10140\AA )
spectrum from MAST taken along the same position as the G430M slit D in order to measure the 
flux ratio of $H\alpha/H\beta$ over the nucleus, and calculated a reddening of $E(B-V) = 0.78$ 
(assuming a galactic reddening curve and intrinsic $H\alpha/H\beta = 2.9$). The reddening is 
significant enough to extinguish portions of the inner NLR, possibly suppressing high-velocity 
dispersion emission from the bicone at $\sim 0$ km s $^{-1}$ and making the double line profile 
more distinct. However, this reddening is not enough to extinguish the AGN broad-line region 
(BLR). As such, it is likely that this dust lane must be additional obscuration external to 
the $''$torus$''$. 

Further evidence for a dust lane can be seen in the blue-shifted southwest lobe from the imaging 
shown in Figure 1. From the end of the absorbing dust lane, an arc of emission is visible along 
the north rim of the lobe $\sim 2''$ from the $``+''$ marking the continuum peak location. 
It remains unclear whether the dust lane crossing across the nucleus 
arcs over the southwest lobe and continues to absorb emission or instead arcs into the ionizing 
bicone and becomes illuminated, similar to Mrk 3 and Mrk 573 \citep{Cre10b,Fis10}. However, both 
situations would help explain the asymmetrical, patchy emission seen in the blue-shifted emission 
lobe. Figure 6 depicts a galactic disk with the north side of the disk coming out of the page, 
which allows for a potential interaction between the disk and ionizing bicone. A full 
understanding of this extended morphology will require more thorough imaging before a more 
accurate claim can be made.

\section{Conclusions}

We have analyzed STIS long-slit G430M spectra of the inner emission knots and nucleus of the 
Seyfert 2 galaxy Mrk 78. We generated kinematic and geometrical models of the NLR and ENLR that 
successfully fit the data, with the final parameters of Mrk 78's outflow bicone given in Table 3. 
While it is not clear at this time how the blue-shifted lobe is interacting with the dust lane 
lying over the nucleus, it is likely that that interaction can explain the dispersed morphology 
present in the given FOC and STIS imaging.

We find that the double peaked emission lines seen in ground-based spectra of Mrk 78 are due to 
an asymmetric distribution of outflowing gas in the NLR. Specifically, the red peak at $+800$ km 
s $^{-1}$ is due to a bright emission-line knot that is redshifted with respect to the systemic 
velocity of the host galaxy. The bright, redshifted knot follows the same flow pattern as the other 
knots, and thus, there is no need to assume that a second AGN resides within Mrk 78. Furthermore, 
all Seyferts that we have modeled to date show outflows \citep{Rui01,Das05,Das06,Fis10}, and thus 
asymmetric distributions of outflowing gas and/or patchy dust extinction maybe a general explanation 
for double-peaked emission-line profiles. With nothing unusual separating Mrk 78 from other AGN, 
it is possible that many double-peaked profiles may be due to outflows in a single AGN as shown 
by our models.

\acknowledgments

\begin{deluxetable}{ccccccc}
\tablecolumns{5}
\footnotesize
\tablecaption{{\it HST}/STIS G430M Observations of Mrk~78}

\tablewidth{0pt}
\tablehead{
\colhead{Slit} & \colhead{Exposure} & \colhead{P.A.}      & \colhead {Offset$^a$} \\
               &    \colhead{(s)}   & \colhead{(degrees)} & \colhead {(arcsec)}
}
\startdata
       A       &      1727          &       88        &        0.125        \\
       B       &      1938          &       88        &       -0.27         \\
       C       &      2052          &       88        &       -0.55         \\
       D       &      1800          &       61        &       -0.05         \\
\enddata
\tablenotetext{a}{Cross-dispersion offset from the optical continuum peak. A positive value
                  corresponds to a displacement south of the slit P.A. axis in Figure 1.}
\end{deluxetable}

\begin{deluxetable}{ccccc}
\tablecolumns{5}
\footnotesize
\tablecaption{{\it HST}/FOC/STIS Imaging Observations of Mrk~78}

\tablewidth{0pt}
\tablehead{
\colhead{Instrument} &    \colhead{Filter}  &    \colhead{Exposure} \\
                     &         &                \colhead{(s)} 
}
\startdata
             STIS    &      MIRVIS        &       120.      \\
             FOC     &      372M          &       896.      \\
             FOC     &      550M          &       1196.     \\
             FOC     &      502M          &       801.      \\
\enddata
\end{deluxetable}

\begin{deluxetable}{ccc}
\tablecolumns{2}
\footnotesize
\tablecaption{Best Fit Model Parameters for Mrk~78$^a$}
\tablewidth{0pt}
\tablehead{
\colhead{Parameter} & \colhead{Values}
}
\startdata
$P.A.$                      &65\deg           \\
$i$                         &30\deg (SW)      \\
$\theta_{max}$              &35\deg           \\
$\theta_{min}$              &10\deg           \\
$v_{max}$                   &1200 km s$^{-1}$ \\
$z_{max}$                   &3200 pc          \\
$r_{t}$                     &700 pc           \\
\enddata
\tablenotetext{a}{The letters in parentheses indicates the side closest to
us.}
\end{deluxetable}

\clearpage

\bibliographystyle{apj}             
\bibliography{apj-jour,paper}       

\clearpage

\figcaption[f1.eps]{Composite of {\it HST} STIS acquisition continuum image (left) with additional 
enlarged [O~III] (F502M filter, as seen in \citet{Sch96}) and UV (F372M filter, mainly [OII] emission) 
FOC images (middle and right respectively). The dust lane sweeping over the AGN is located at a position 
angle of $\sim$123\deg ~southeast to northwest. STIS slit positions have been superimposed and the 
red-shifted emission knot responsible for the secondary radial velocity peak is circled on the [O~III] image.}

\figcaption[f2.eps]{Summation of spectral flux from slits A, B, and C as a function of radial 
velocity for the [O~III] $\lambda$5007 emission line with respect to the systemic velocity. 
Double peaked [O~III] $\lambda$4959 emission-line profile is seen to the left.}

\figcaption[f3.eps]{Radial velocities (top), FWHM (middle), and normalized total flux (bottom) 
of [O~III] measurements for slits A and B. The bracket in slit B shows the location of the high 
flux cloud responsible for the secondary radial velocity peak.}

\figcaption[f4.eps]{Radial velocities (top), FWHM (middle), and normalized total flux (bottom) 
of [O~III] measurements for slits C and D.}

\figcaption[f5.eps]{Shaded regions represent the kinematic model chosen as the best fit for our 
radial velocity data set for all four slits. Parameters used to create this model are given in 
Table 1. Lighter regions represent kinematics for the side of each cone near parallel to our line 
of sight, darker regions represent kinematics for the side near perpendicular to our line of 
sight.}

\figcaption[f6.eps]{Geometric model of the NLR based on parameters from Table 1, shown as viewed 
from Earth. Inner disk geometry is taken from \citet{Kin00} depicting the top of the disk 
coming out of the sky. The yellow bar represents the position angle of the dust lane traveling 
over the nucleus. The bright redshifted knot is in the northeast portion of the bicone.}

\figcaption[f7.eps]{Same as Figure 6, except the inner disk geometry depicts the bottom of the 
disk coming out of the sky. The yellow bar represents the position angle of the dust lane 
traveling over the nucleus.}

\clearpage

\begin{figure}
\plotone{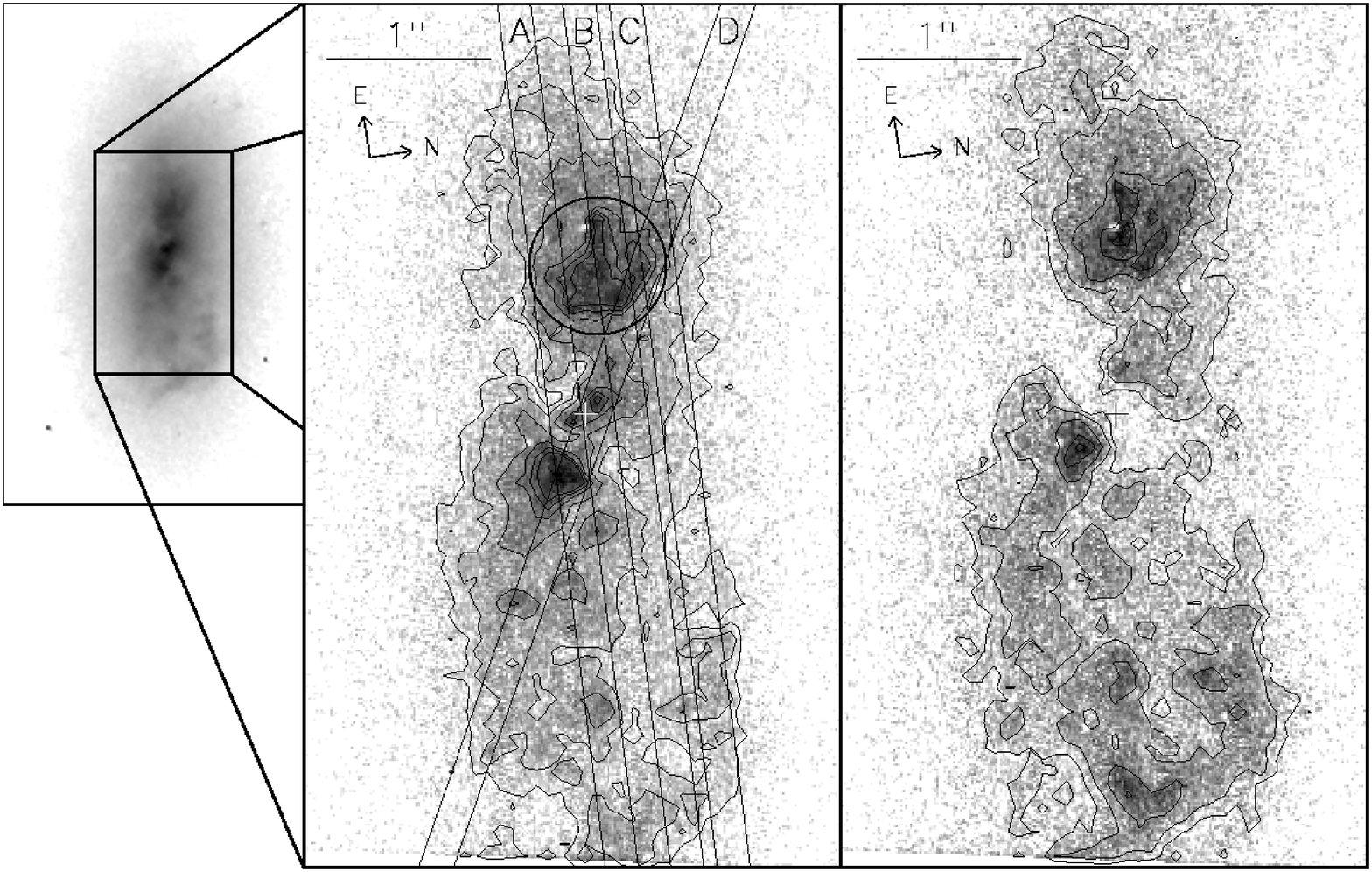}
\\Fig.~1.
\end{figure}

\begin{figure}
\plotone{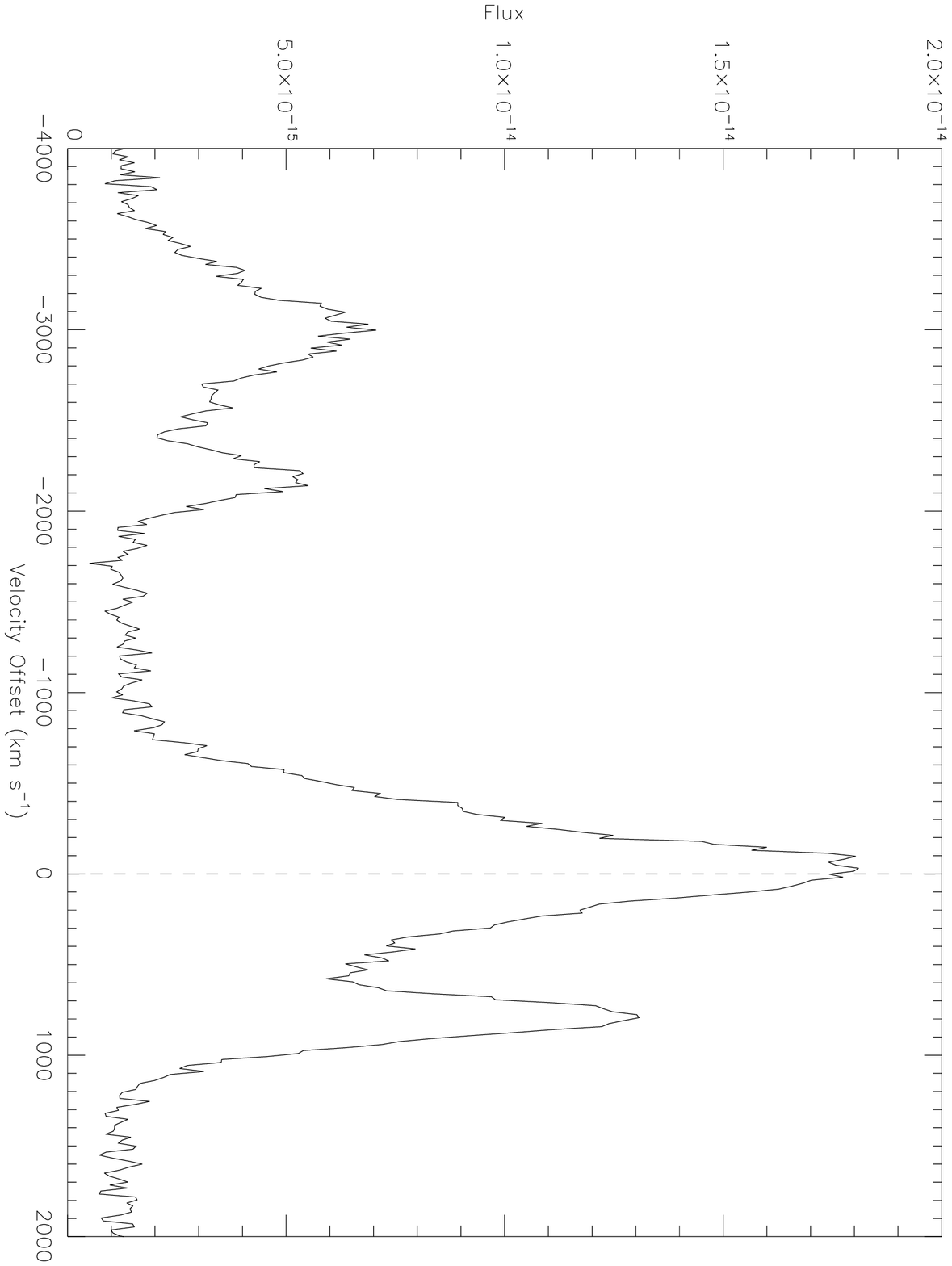}
\\Fig.~2.
\end{figure}

\begin{figure}
\plotone{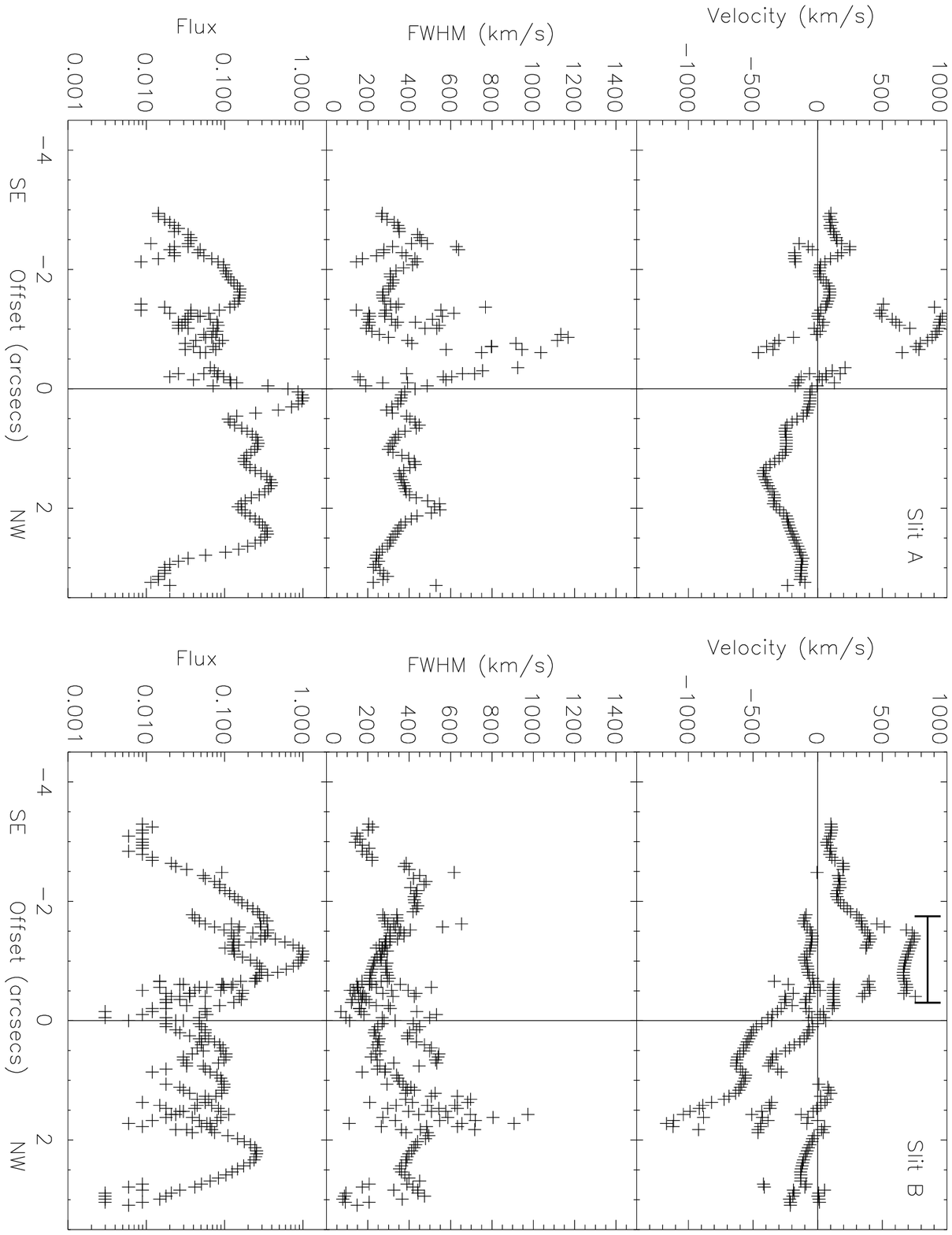}
\\Fig.~3.
\end{figure}

\begin{figure}
\plotone{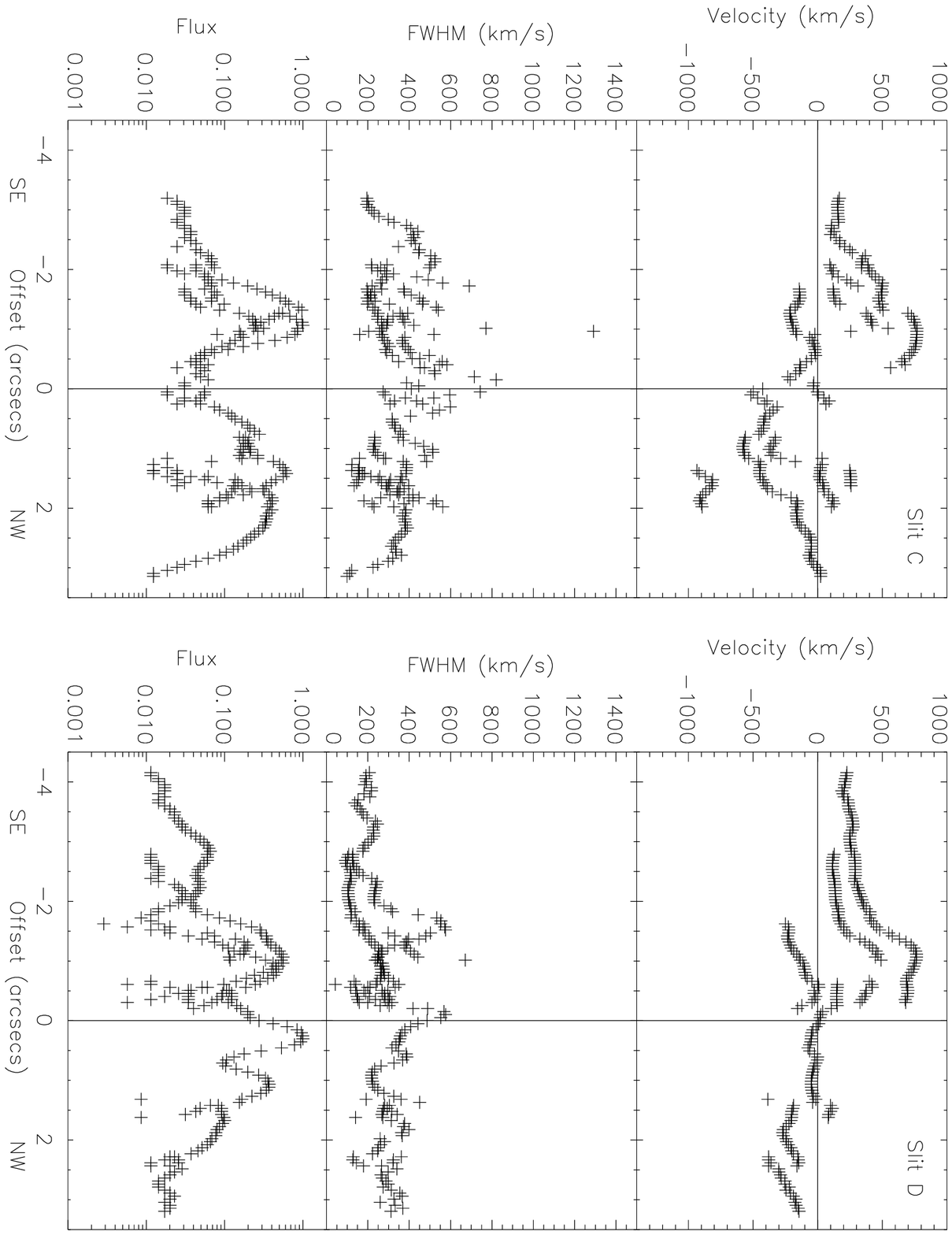}
\\Fig.~4.
\end{figure}

\begin{figure}
\plotone{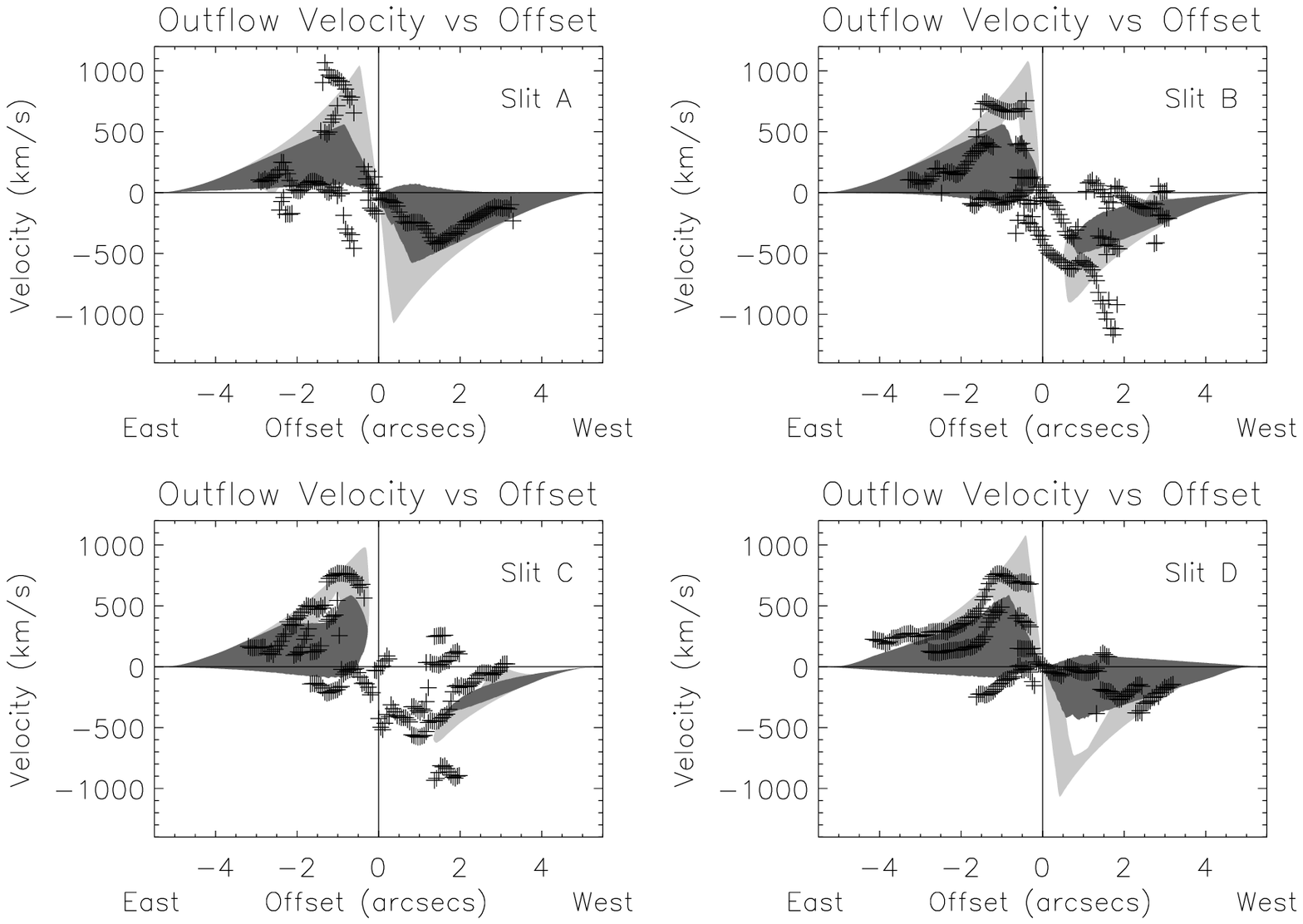}
\\Fig.~5.
\end{figure}

\begin{figure}
\plotone{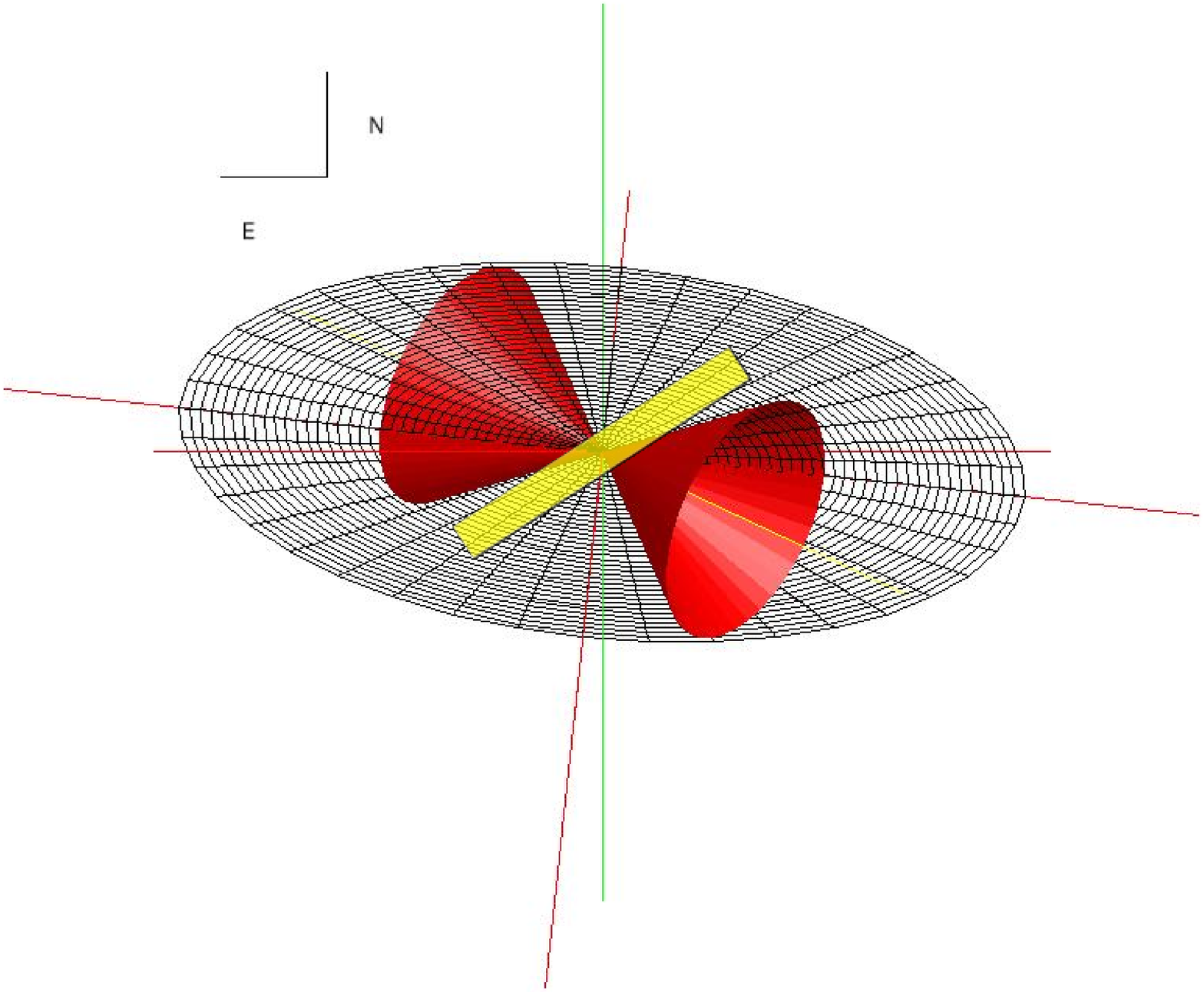}
\\Fig.~6.
\end{figure}

\begin{figure}
\plotone{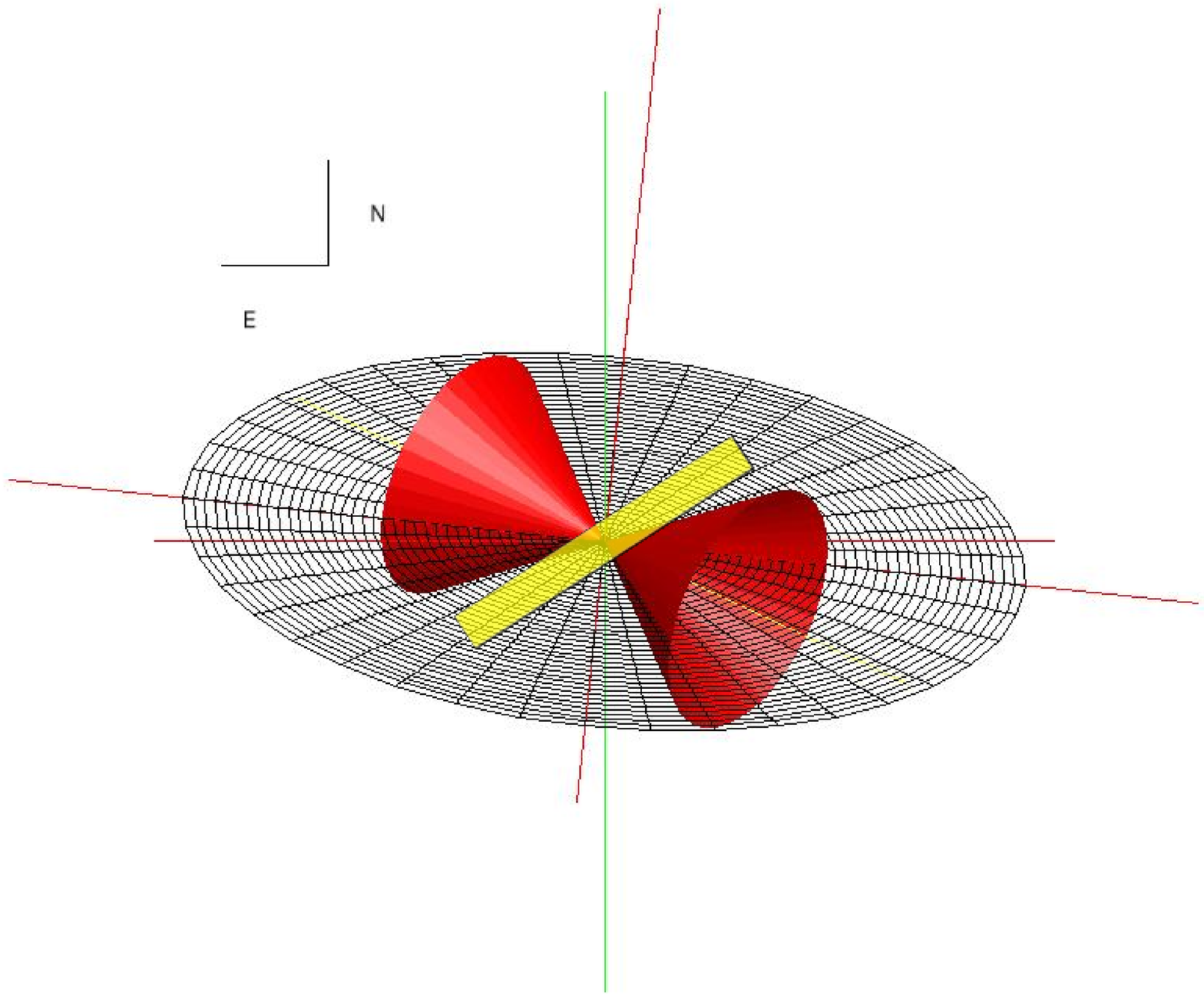}
\\Fig.~7.
\end{figure}
\end{document}